\documentclass[12pt]{article}
\usepackage[english]{babel}
\usepackage{graphicx}
\usepackage{amssymb}
\usepackage{amsfonts}
\usepackage{amsmath}
\usepackage{bm}
\usepackage[a4paper,margin=20mm,nohead]{geometry}

\newcommand{\BB}{\bm{B}}
\newcommand{\rr}{\bm{r}}
\newcommand{\pp}{\bm{p}}
\newcommand{\AAA}{\bm{A}}
\newcommand{\aaa}{\bm{a}}
\newcommand{\vq}{\bm{q}}
\newcommand{\HH}{{\cal H}}
\newcommand{\DD}{{\cal D}}
\newcommand{\RR}{{\mathbb R}}
\newcommand{\cQ}{{\cal Q}}
\newcommand{\ZZ}{{\mathbb Z}}
\newcommand{\NNN}{{\mathbb N}}
\newcommand{\black}{\rule{7pt}{7pt}}
\newcommand{\eps}{{\varepsilon}}
\newcommand{\TT}{{\mathbb T}}
\newcommand{\rot}{{\rm rot\,}}
\newcommand{\w}{\omega}  
\newcommand{\GtB}{G_3(\vq,\vq';\zeta|\omega,B)}
\newcommand{\Gt}{G_3(\vq,\vq';\zeta|\omega')}
\newcommand{\Qt}{Q_3(\zeta;\vq|\w')}
\newcommand{\QtB}{Q_3(\zeta;\vq|\w,B)}
\newcommand{\QtW}{Q_3(\zeta;\vq|\Omega/2,\w)}
\newcommand{\psid}{\psi_{mn}^{(2)}}
\newcommand{\EdB}{E_{mn}^{(2)}(\w,B)}
\newcommand{\Ed}{E_{mn}^{(2)}(\Omega/2)}
\newcommand{\EdW}{E_{mn}^{(2)}(\Omega/2)}
\newcommand{\Goq}{G_1(q_\|,q_\|;\zeta-\Ed|\Omega/2)}
\newcommand{\Htw}{H^{(3)}(\omega)}
\newcommand{\Ho}{H^{(1)}(\omega)}
\newcommand{\HtB}{H^{(3)}(\omega,B)}
\newcommand{\StB}{S_3(\vq,\vq'|\omega,B)}
\newcommand{\Gto}{G_3(\vq,\vq';\zeta|\omega')}
\newcommand{\Ht}{H^{(3)}(\omega')}
\newcommand{\St}{S_3(\vq,\vq'|\omega')}
\newcommand{\HdB}{H^{(2)}(\omega,B)}
\newcommand{\Hdw}{H^{(2)}(\omega)}
\newcommand{\HdW}{H^{(2)}(\Omega/2)}
\newcommand{\QtBo}{Q_3(\zeta;\bm{0}|\w,B)}
\newcommand{\limsum}{\sum\limits}
\newcommand{\GoBq}{G_1(q_\|,q_\|;\zeta-\EdB|\omega)}

\begin{document}

\title{Spectrum structure for a
three-dimensional periodic array of quantum dots in a uniform magnetic field}

\author{J.~Br\"uning$^{a}\!$, V.~V.~Demidov$^{b}\!$, V.~A.~Geyler$^{a,b}\!$,
A.~V.~Popov$^{c}\!$}

\maketitle

\begin{quote}
{\small {\em a) Institut f\"ur Mathematik, Humboldt Universit\"at zu Berlin,\\
\phantom{a) } Rudower Chaussee 25, 12489 Berlin, Germany\\
b) Department of Mathematical Analysis, Mordovian State University\\
\phantom{b) } 430000 Saransk, Russia\\
c) Department of Mathematical Methods of Programming, Mordovian State University\\
\phantom{c) } 430000 Saransk, Russia;}\\
\phantom{a) }\texttt{bruening@mathematik.hu-berlin.de}, \texttt{geyler@mrsu.ru}}
\end{quote}

\medskip

\begin{abstract}
\noindent By means of the operator extension theory, we construct
an explicitly solvable model of a simple-cubic three-dimensional
regimented array of quantum dots in the presence of a uniform
magnetic field. The spectral properties of the model are studied.
It is proved that for each magnetic flux the band is the image of
the spectrum of the tight-binding operator under an analytical
transformation. In the case of rational magnetic flux the spectrum
is described analytically. The flux-energy and angle-energy
diagrams are obtained numerically.
\end{abstract}

\section{Introduction}

The problem of commensurability in periodic systems is a source of
fascination in various branches of physics and is related to the
appearance of the fractal spectrum for such systems.
M.~Ya.~Azbel', D.~R.~Hofstadter, and
G.~H.~Wannier~\cite{Azb}--\cite{Wan} were the first scientists who
studied the simplest model of the two-dimensional (2D) electron
system in a uniform magnetic field and discovered a self-similar
energy spectrum, called ``Hofstadter's butterfly''. This picture
shows the dependence of the energy on the magnetic field for
two-dimensional periodic system in the tight-binding approximation
and confirms the conjecture that the Hamiltonian of this model
(which is reduced to the so-called {\it Harper operator}) has the
Cantor spectrum at irrational value of the magnetic flux (the
famous ``Ten Martini Problem''). Very recently this problem has
been solved affirmatively~\cite{AvKr}, \cite{AvJi}.

It was assumed the long time that in the three-dimensional (3D)
case the gaps in the energy spectrum, in general, are closed and
the fractal structure of the spectrum is impossible. Recently
M.~Koshino, H.~Aoki and co-authors have considered the dependence
of the energy spectrum of a 3D periodic quantum system on the
orientation of the magnetic field and on the strength of the field
\cite{KAKKO}--\cite{KA}. Using semiclassical methods they have
reduced the Schr\"odinger equation with the 3D tight-binding
Hamiltonian to the 1D Harper equation and shown that in the
obtained angle-energy and flux-energy diagrams a fractal structure
appears. Note the tight-binding method is based on a series of
approximations of the initial Schr\"odinger equation that can
change considerably the structure of the spectrum, e.g., the
tight-binding Hamiltonian for a single-atomic lattice has only one
magnetic band in the spectrum.

Another explicitly solvable model of 3D periodic quantum systems
was presented in~\cite{GD}--\cite{BDG2}. We have considered
spectral properties of the 3D Landau operator perturbed by
periodic point potential. It was shown numerically in~\cite{BDG1},
\cite{BDG2} that the angle-energy and flux-energy diagrams for
different bands of the spectrum have a fractal structure even in
case of cubic lattice. Note that the dispersion equation in this
model has a transcendental form and the analytical study of the
spectrum becomes very difficult.

In this paper we consider the quantum-mechanical model for the 3D
periodic array of quantum dots obtained by means of operator
extension theory~\cite{Pav1}--\cite{GPPMod}. It may be noted that
this method was successfully applied for construction and spectral
investigation of quantum periodic systems in two dimensions in our
previous works~\cite{GP}, \cite{GPPO}. In particular, the model
allows, in contrary to the tight-binding approximation method, to
describe all bands of the spectrum. On the other hand, the
dispersion equation of our model has more simple form than the
corresponding one for the 3D Landau operator perturbed by periodic
point potential, and {\it in the case of rational magnetic flux}
the spectrum of 3D periodic array of quantum dots is described
analytically.

In the framework of our model, each Fock--Darwin level of a single quantum dot
broads into magnetic band which in its turn splits into subbands. This is
stipulated by the spectrum structure of irreducible representations of the
magnetic translation group (for the first time it was shown by
J.~Zak~\cite{Zak}). If the numbers of the flux quanta of the magnetic field
through each face of elementary cell of the Bravais lattice $\Lambda$ of the
system are rational (in this case the magnetic field is said to be rational
with respect to the lattice $\Lambda$), then there is a finite number of
subbands for a fixed magnetic band. If the fluxes through adjacent faces of
elementary cell of the Bravais lattice of the system are equal to an
irreducible fractions $N_1/M_1$, $N_2/M_2$ and $N_3/M_3$, then the numbers of
subbands are defined by the numbers $M_1$, $M_2$ and $M_3$. Therefore the
magnetic band arising from the fixed Fock--Darwin level can transform to a
Cantor set when one of these fluxes tends to an irrational number. For the
simple-cubic array, we prove that in both cases of rational and irrational flux
the band is the image of the spectrum of the corresponding tight-binding
operator under an analytic function. Hence, a fractal structure of angle-energy
and flux-energy diagrams for 3D regimented array of quantum dots inherits from
such diagrams for the tight-binding operator. It should be noted that similar
description of the spectrum was obtained in the 2D case in \cite{GPPMod},
\cite{GP} and for quantum graphs in \cite{Pan} and \cite{BGP}.

Our model is directly relevant to the so-called 3D regimented
arrays of semiconductor quantum dots which attract the increased
attention lately~\cite{Redl}--\cite{Bimberg}. The regimentation
along all three directions results in the formation of an
artificial crystal, where quantum dots play the role of
atoms~\cite{LB}, \cite{LB2}. O.~L.~Lazarenkova and A.~A.~Balandin
in~\cite{LB} have analyzed the band structure of such systems
using envelope function approximation.

\section{The description of the model}

For construction of our model we start with the quantum-mechanical Hamiltonian $H_d$ of
a single 3D quantum dot subjected to a uniform magnetic field $\bm{B}$:
\begin{equation}
H_d=H_0+V(\rr),
\end{equation}
where $H_0$ is the Hamiltonian of a single electron in the magnetic field,
\begin{equation}
H_0=\frac{1}{2m^*}\left(i\hbar\bm{\nabla}+\frac{e}{c}\AAA(\rr)\right)^2.
\end{equation}
Here $e$ and $m^*$ are the charge and the effective mass of the
electron respectively, and $\AAA(\rr)$ is the vector potential of
the field $\BB$ (i.e. $\BB=\rot\AAA$), $V(\rr)$ is the confining
potential of a three-dimensional quantum dot. In our model we
choose $V(\rr)$ as the potential of a spherical quantum well:
\begin{equation}
V(\rr)=V(x,y,z)=\frac{m^*\omega^2}{2}\left(x^2+y^2+z^2\right).
\end{equation}
Here the characteristic sizes $d_x$, $d_y$ and $d_z$ of the
quantum well are related to the frequency $\omega$ as follows:
$$
d_x,d_y,d_z\sim\sqrt{\hbar/(m^*\omega)}.
$$

Now let us consider a periodic three-dimensional array of such quantum dots. We
suppose that centers of the quantum dots coincide with nodes of simple cubic
crystallic lattice $\Lambda$ with the basis $(\aaa_1,\aaa_2,\aaa_3)$ (Fig. 1).\\


The state space of our model is the direct sum of the state spaces $L^2(\RR^3)$
of single quantum dots:
\begin{equation}
\HH=\sum\limits_{\bm{\lambda}\in\Lambda}{}^\oplus \HH_{\bm{\lambda}},\quad
\HH_{\bm{\lambda}}=L^2(\RR^3)\ \mbox{ for each }\ \bm{\lambda}\in\Lambda.
\end{equation}
The following direct sum
\begin{equation}
H^0=\sum\limits_{\bm{\lambda}\in\Lambda}{}^\oplus H_\lambda, \;
H_{\bm{\lambda}}=H_d\;\mbox{for each }\bm{\lambda}\in\Lambda,
\end{equation}
is the unperturbed Hamiltonian of the model. This operator is the
Hamiltonian of a set of isolated quantum dots. To take into
account the charge carrier tunnelling between dots we use the
"restriction -- extension"\ procedure of the operator extension
theory \cite{Pav1}, \cite{Pav2}. We denote by $\DD$ the set of
functions $f$ from the domain $\DD(H_d)$ each of which vanishes in
a neighborhood of zero. Let $S_d$ be the restriction of the
operator $H_d$ to the set $\DD$ and let $S$ is the following
direct sum
\begin{equation}
S=\sum\limits_{\bm{\lambda}\in\Lambda}{}^\oplus S_{\bm{\lambda}},\quad
S_{\bm{\lambda}}=S_d\ \mbox{ for each }\ \bm{\lambda}\in\Lambda.
\end{equation}
We seek the "true"\ Hamiltonian $H$ of the quantum dots array among non-trivial
self-adjoint extensions of the operator $S$ such as
$\DD(H)\cap\DD(H^0)=\DD(S)$.

The resolvent $R$ of the operator $H$ and the resolvent $R^0$ of the operator
$H^0$ are related with the Krein resolvent formula (see, e.g. \cite{GPPMod}):
\begin{equation}
                \label{Krein}
R(\zeta)\equiv R_A(\zeta)=R^0(\zeta)-g(\zeta)[Q(\zeta)+A]^{-1}g^*(\bar \zeta),
\end{equation}
where operator valued holomorphic functions $g(\zeta)$ and
$Q(\zeta)$ are the so-called Krein $\Gamma$- and $\cQ$-functions
of the operator $H^0$ respectively, and $A$ is a self-adjoint
operator in the deficiency space $l^2(\Gamma)$ (the so-called
"interaction"\ operator). The main part in the study of the
spectrum is played by the $\cQ-$function $Q(\zeta)$ and operator
$A$.

\subsection{The $\cQ$-function of the model}

In the considered case the $\cQ$-function $Q(z)$ has the diagonal matrix in the
space $l^2(\Lambda)$:
\begin{equation}
Q(\lambda,\lambda';z)=\delta_{\lambda\lambda'}q(z),\;\;
\lambda,\lambda'\in\Lambda,
\end{equation}
where $q(z)$ is the $\cQ$-function of the operator $H_d$. To
obtain the function $q(z)$ we find preliminarily a general
expression for the $\cQ$-function $\QtB$ of the operator $H_d$ at
an arbitrary point $\vq\in\RR^3$.

We denote by $\Htw$ the Hamiltonian of spherical quantum dot without magnetic
field:
\begin{equation}
\Htw=-\frac{\hbar^2}{2m^*}\Delta+\frac{m^*\w^2}{2}(x_1^2+x_2^2+x_3^2).
\end{equation}
Let $\GtB$ be the Green function of the operator $H_d\equiv\HtB$, $\Gto$ be the
Green function of the operator $\Ht$, and $\Qt$ be the $\cQ$-function of the
operator $\Ht$. We denote by $\St$ and $\StB$ the singularities of the
functions $\Gt$ and $\GtB$ at a point $\vq$ respectively. It is known that for
each $\w$ and $\w'$ the singularities $\St$ and $\StB$ coincide and have the
following form (see, e.g. \cite{BGL}):
\begin{equation}
\label{e11} \St=\StB=\frac{m^*}{2\pi\hbar^2}\frac{1}{|\vq-\vq'|}.
\end{equation}
It implies for each $\w'$ the following form of the function $\QtB$:
$$
\QtB=\lim\limits_{\vq\to\vq'}[\Gt-\St]=
$$
\begin{equation}
\label{e8} =\lim\limits_{\vq\to\vq'}[\GtB-\Gt]+\Qt.
\end{equation}
We denote by $\HdB$ the Hamiltonian of two-dimensional symmetric quantum dot
subjected with magnetic field $\BB$ (the field $\BB$ is perpendicular to the
quantum dot plane):
\begin{equation}
\HdB=\frac{1}{2m^*}\left(-i\hbar\nabla-\frac{e}{c}\AAA(\rr)\right)^2+
\frac{m^*\w^2}{2}(x_1^2+x_2^2).
\end{equation}
and by $\Hdw$ the Hamiltonian of the same dot without magnetic field:
\begin{equation}
\Hdw=-\frac{\hbar^2}{2m^*}\Delta+\frac{m^*\w^2}{2}(x_1^2+x_2^2).
\end{equation}
We use the standard notations $\omega_c=|e\BB|/(cm^*)$ for the
cyclotron frequency and $\Omega=\sqrt{\omega_c^2+4\omega^2}$ for
the hybrid frequency. It is easy to show that the eigenfunctions
of the operators $\HdB$ and $\HdW$ coincide and have the following
form in the polar coordinates $(\rho,\phi)$:
$$
\psid(\rho,\phi)=\frac{1}{\sqrt{2\pi}l_\Omega^{1+|m|}}
\sqrt{\frac{n!}{2^m(n+|m|)!}}\rho^{|m|}\exp
\left(-\frac{\rho^2}{4\l_\Omega^2}\right)
L_n^{|m|}\left(\frac{\rho^2}{2\l_\Omega^2}\right)\exp (im\phi),
$$
\begin{equation}
m\in\ZZ, n=0,1,2,\dots,
\end{equation}
where $l_\Omega=\sqrt{\hbar/(m^*\Omega)}$, $L_n^{|m|}$ is the Laguerre-Sonin
polynomial of the power $n$. Corresponding eigenvalues $\EdB$ and $\Ed$ of the
operators $\HdB$ and $\HdW$ have the following form:
\begin{equation}
\label{e15} \EdB=\frac{\hbar\w_c}{2}m+\frac{\hbar\Omega}{2}(2n+|m|+1),
\end{equation}
\begin{equation}
\Ed=\EdB-\frac{\hbar\w_c}{2}m=\frac{\hbar\Omega}{2}(2n+|m|+1).
\end{equation}
Further we shall denote coordinates $(\rho,\phi,z)$ of a point $\vq$ by
$(\vq_\bot,q_\|)$, where $\vq_\bot=(\rho,\phi)$, $q_\|=z$. We have the
following representation of the function $\QtB$:
$$
\QtB=\sum\limits_{m=-\infty}^\infty\limsum_{n=0}^\infty
|\psid(\vq_\bot)|^2[\GoBq-
$$
\begin{equation}
-\Goq]+\QtW,
\end{equation}
where $G_1(x,x';\zeta|\w)$ is the Green function of the one-dimensional
harmonic oscillator $\Ho$:
\begin{equation}
\Ho=-\frac{\hbar^2}{2m^*}\frac{d^2}{dx^2}+\frac{m^*\w^2}{2}x^2.
\end{equation}
It is known that the function $G_1(x,x';\zeta|\w)$ has the following form (see,
e.g. \cite{GCh}):
$$
G_1(x,x';\zeta|\w)=\frac{1}{\sqrt{\pi}\hbar\w l_\w}\Gamma\left(\frac{1}{2}-
\frac{\zeta}{\hbar\w}\right)U\left(-\frac{\zeta}{\hbar\w},
\frac{\sqrt{2}\max(x,x')}{l_\w}\right)\times
$$
\begin{equation}
\times U\left(-\frac{\zeta}{\hbar\w}, -\frac{\sqrt{2}\min(x,x')}{l_\w}\right).
\end{equation}
Therefore we obtain the following expression for the function $\QtB$:
$$
\QtB=\frac{1}{2\pi^{3/2}\hbar l_\Omega^2}\exp\left(
-\frac{\vq_\bot^2}{l_\Omega^2}\right)\limsum_{m=-\infty}^\infty
\left(\frac{\vq_\bot^2}{2l_\Omega^2}\right)^{|m|}\limsum_{n=0}^\infty
\frac{n!}{(n+|m|)!}\times
$$
$$
\times\left[L_n^{|m|}\left(\frac{\vq_\bot^2}{2l_\Omega^2}\right)\right]^2
\left(\frac{\Gamma\left(\frac{1}{2}+\frac{\EdB-\zeta}{\hbar\w}\right)} {\w
l_{\w}}
U\left(\frac{\EdB-\zeta}{\hbar\w},\frac{\sqrt{2}q_\|}{l_{\w}}\right)\times
\right.
$$
$$
\times U\left(\frac{\EdB-\zeta}{\hbar\w},-\frac{\sqrt{2}q_\|}{l_{\w}}\right)-
\frac{\sqrt{2}\Gamma\left(\frac{1}{2}+\frac{2(\EdW-\zeta)}{\hbar\Omega}\right)}
{\Omega l_\Omega}\times
$$
\begin{equation}
\label{e25} \left. \times
U\left(\frac{2(\EdW-\zeta)}{\hbar\Omega},\frac{q_\|}{l_\Omega}\right)
U\left(\frac{2(\EdW-\zeta)}{\hbar\Omega},-\frac{q_\|}{l_\Omega}\right)
\right)+\QtW,
\end{equation}
Now we turn to the desired $\cQ$-function $q(z)=\QtBo$. It is known that for
each $\w$ the function $Q_3(\zeta;0|\w)$ has the following form (see, e.g.
\cite{BGL}):
$$
Q_3(\zeta;0|\w)\equiv
q_0(z)=-\frac{m^*}{\pi\hbar^2l_\w}\frac{\Gamma\left(\frac{3}{4}-
\frac{\zeta}{2\hbar\w}\right)}{\Gamma\left(\frac{1}{4}-
\frac{\zeta}{2\hbar\w}\right)},
$$
Then it is easy to show that the $q(z)$ has the following form:
\begin{equation}
q(z)=\frac{1}{4\pi\hbar l_\Omega^2}\sum\limits_{n=0}^\infty\left\{
\frac{1}{\omega l_{\omega}}\cdot\frac{\Gamma\left(\frac{1}{4}+
\frac{\Omega}{2\omega}\left(n+\frac{1}{2}\right)-\frac{z}{2\hbar\omega}
\right)}
{\Gamma\left(\frac{3}{4}+\frac{\Omega}{2\omega}\left(n+\frac{1}{2}\right)-
\frac{z}{2\hbar\omega}\right)}- \frac{\sqrt{2}}{\Omega
l_\Omega}\cdot\frac{\Gamma\left(\frac{3}{4}+n- \frac{z}{\hbar\Omega}\right)}
{\Gamma\left(\frac{5}{4}+n-\frac{z}{\hbar\Omega}\right)} \right\}+q_0(z).
\end{equation}

\subsection{The "interaction" operator of the model}

The matrix
$(A(\bm{\lambda},\bm{\lambda}'))_{\bm{\lambda},\bm{\lambda}'\in\Lambda}$ of the
"interaction" operator $A$ in our model is described by the following conditions
(cf. \cite{GPPMod}):\\
1) $A$ is the bounded self-adjoint operator in $l^2(\Lambda)$, i.e.
$A(\bm{\lambda},\bm{\lambda}')=\overline{A(\bm{\lambda}',\bm{\lambda})}$ for
each $\bm{\lambda},\bm{\lambda}'\in\Lambda$;\\
2) $A$ is the nearest-neighbour-interaction operator, i.e.
$A(\bm{\lambda},\bm{\lambda}')=0$, if $|\bm{\lambda}-\bm{\lambda}'|>\inf
\{|\bm{\lambda}-\bm{\lambda}'|\,:\,\bm{\lambda},\bm{\lambda}'\in\Lambda,
\bm{\lambda}\neq\bm{\lambda}'\}$;\\
3) $A$ is invariant with respect to a natural unitary
representation of the magnetic translation group in the space
$l^2(\Lambda)$, i.e.
\begin{equation}
A(\bm{\lambda},\bm{\lambda}+\bm{\mu})=\exp [\pi
i\bm{\xi}\cdot(\bm{\mu}\times\bm{\lambda})]A(0,\bm{\mu}),\;\;
\bm{\lambda},\bm{\mu}\in\Lambda,
\end{equation}
where $\bm{\xi}=\BB/\Phi_0$ is the vector of magnetic flux density (here
$\Phi_0=2\pi\hbar c/e$ is the magnetic field quantum).

Therefore in the case of simple cubic crystallic lattice $\Lambda$ the matrix
of the operator $A$ is fully determined only by the elements
$A(\bm{\lambda},0)$, $\bm{\lambda}\in\Lambda$:
\begin{equation}
A(\bm{\lambda},0)=
\begin{cases}
\alpha_1,\,\bm{\lambda}=\pm\aaa_1,\cr
       \alpha_2,\,\bm{\lambda}=\pm\aaa_2,\cr
       \alpha_3,\,\bm{\lambda}=\pm\aaa_3,\cr
       0,\,\mbox{in the other cases}.
\end{cases}
\end{equation}
Here $\alpha_1,\alpha_2$ and $\alpha_3$ are some real numbers (the so-called
"coupling constants").

The construction of the Hamiltonian $H\equiv H_A$ of 3D periodic quantum dots
array in a uniform magnetic field is complete.

\section{The structure of the spectrum of the Hamiltonian $H_A$}

In the framework of our model it is possible to show that the spectrum of the model
Hamiltonian $H_A$ consists of the pure point and the band parts. The description of each
band of the spectrum is mathematically reduced (without any additional simplifications
or approximations) to the investigation of the only energy band in the tight-binding
model. Moreover in the case of so-called rational magnetic flux the structure of the
bands may be described analytically.

To investigate the spectrum of the operator $H_A$ we start with
the spectrum of the Hamiltonian $H_d$ of single quantum dot. The
spectrum $\sigma(H_d)$ is discrete and consists of the eigenvalues
$E_{mn_\rho n_z}$ (the Fock--Darwin levels),
$$
E_{mn_\rho n_z}=\frac{\omega_c}{2}m+\frac{\Omega}{2}
(2n_\rho+|m|+1)+\frac{\omega}{2}(2n_z+1),
$$
\begin{equation}
m\in\ZZ,\quad n_\rho,n_z\in\NNN\cup\{0\}.
\end{equation}
Thus, the spectrum of the Hamiltonian $H^0$ (the Hamiltonian of a set of isolated
quantum dots) is pure point one and consists of the same points $E_{mn_\rho n_z}$, and
each point $E_{mn_\rho n_z}$ is infinitely degenerated in the spectrum of
$H^0$. The Krein formula implies the following proposition.\\[2mm]
{\bf Proposition 1.} {\it Let a real number $E$ be not a point of
the spectrum of $H^0$, then the following assertions are
equivalent:
\begin{itemize}
\item[1)] $E\in\sigma(H_A)$; \item[2)] the operator $q(E)I+A$ is
not continuously invertible in the space $l^2(\Lambda)$. \black
\end{itemize}
}

Therefore, to describe the spectrum of the Hamiltonian $H_A$ we
can consider the spectrum of the operator $A$ and the behavior of
the function $q(x)$ on the real axis.

We denote by $H_{TB}$ the standard  tight-binding Hamiltonian:
\begin{equation}
\label{HTB}
H_{TB}=-\sum\limits_{\bm{\lambda},\bm{\lambda}'}t_{\bm{\lambda},\bm{\lambda}'}
\exp(i\theta_{\bm{\lambda},\bm{\lambda}'})C_{\bm{\lambda}}^+C_{\bm{\lambda}'}^{\phantom{+}},
\end{equation}
where the summation is over nearest-neighbor nodes of $\Lambda$,
$C_{\bm{\lambda}}^+$ and $C_{\bm{\lambda}}^{\phantom{+}}$ are the
standard creation and annihilation operators at the node
$\bm{\lambda}$,
$\theta_{\bm{\lambda}\bm{\lambda}'}=-\dfrac{e}{\hbar
c}\int\limits_{\bm{\lambda}}^{\bm{\lambda}'} \AAA \cdot d\bm{l}$
is the Peierls phase factor arising from the magnetic flux. It is
easy to show that the interaction operator $A$ coincides with the
operator $H_{TB}$ with
$t_{\bm{\lambda},\bm{\lambda}'}=-A(\bm{\lambda},\bm{\lambda}')$
for each $\bm{\lambda},\bm{\lambda}'\in\Lambda$.

As for the function $q$, it is known that for real $x$ the function $q(x)$
monotone increases from $-\infty$ to $+\infty$ on each of the intervals
$(-\infty,\eps_0)$, $(\eps_0,\eps_1)$, $(\eps_1,\eps_2)$, \ldots, where the
numbers $\eps_0,\eps_1,\ldots$ are the eigenvalues $E_{mn_\rho n_z}$ putting in
ascending order (Fig.~2):
\begin{equation}
E_{000}=\eps_0<\eps_1<\ldots<\eps_n<\ldots.
\end{equation}


Hence, there exists the multivalued real-analytical inverse function $\chi(x)$
($x\in\RR$) having continuos branches $\chi_n(x)$ ($n=0,1,\ldots$) with the
values in the intervals $(-\infty,\eps_0)$, $(\eps_0,\eps_1)$,\ldots (Fig.~3).
Let us denote by $\tau_n$ the function $\tau_n(x)=-\chi_n(x)$.


The following theorem
describes the structure of the spectrum of the Hamiltonian $H_A$.\\[2mm]
{\bf Theorem 1.} {\it The spectrum $\sigma(H_A)$ of the Hamiltonian $H_A$ is the union
of two parts, $\sigma(H_A)=\Sigma_1\bigcup\Sigma_2$. The part $\Sigma_1$ consists of
infinitely degenerated eigenvalues $E_{mn_\rho n_z}$, where $m\neq0$. The band part
$\Sigma_2$ of the spectrum of the operator $H_A$ has the following form:
$\Sigma_2=\bigcup\limits_{n=0}^{\infty}Y_n$, where the set $Y_n$ is the image of the
spectrum $\sigma(H_{TB})$ of the tight-binding operator $H_{TB}$ under the
real-analytical function $\tau_n$} \black

Therefore the every band $Y_n$ is a result of deformation of a
unique energy band in the tight-binding model under the
corresponding function $\tau_n$. This fact explains, in
particular, various form of the spectrum in different energy
bands.

To continue the study of the spectrum of the operator $H_A$ we
consider the case of a rational field $\BB$. In this case the part
$\Sigma_2$ of the spectrum $\sigma(H_A)$ may be described
analytically.

Fix a basis $(\aaa_1,\aaa_2,\aaa_3)$ in $\Lambda$ and denote
$$
\eta_1=\frac{\BB(\aaa_2\times\aaa_3)}{\Phi_0},\quad
\eta_2=\frac{\BB(\aaa_3\times\aaa_1)}{\Phi_0},\;
\eta_3=\frac{\BB(\aaa_1\times\aaa_2)}{\Phi_0};
$$
by the supposition, the numbers $\eta_j$ are rational. Therefore,
a basis $(\aaa'_1,\aaa'_2,\aaa'_3)$ can be chosen in such a way
that $\eta_1=\eta_2=0$,
$$
\eta_3\equiv\eta=\frac{\BB(\aaa'_1\times\aaa'_2)}{\Phi_0}=\frac{N}{M}>0,
$$
where $N$ and $M$ are coprime positive integers.

Then by means of harmonic analysis on the magnetic translation group we reduce
the spectral problem for the operator $H_A$ to the eigenvalue problem of linear
algebra. Namely, to define the band part of the spectrum of $H_A$ we have to
solve the following dispersion equations:
\begin{equation}
\label{DispEq}
\det[q(E)+\widetilde A(\pp)]=0,
\end{equation}
where $\pp=(p_1,p_2,p_3)\in\TT_3=[\,0,M^{-1})\times[\,0,1)\times[\,0,1)$ is the
quasimomentum, $\widetilde A(\pp)$ is the $M\times M$ matrix with the following
elements:
$$
\widetilde A(m_1,m_2;\pp)=\sum\limits_{\lambda_1,\lambda_2,\lambda_3\in\ZZ}
A(\lambda_1\aaa'_1+(\lambda_2M+m_1-m_2)\aaa'_2+\lambda_3\aaa'_3,\bm{0})\times
$$
\begin{equation}
\label{Ap}
\exp\left[-2\pi i\left(\lambda_1p_1+\lambda_2p_2+\lambda_3p_3+\frac{\eta}{2}
\lambda_1(M\lambda_2+m_1+m_2)\right)\right].
\end{equation}
It is obvious that for each $\pp$ the series~(\ref{Ap}) is the finite sum.

The following theorem describes the spectrum of the operator $H_A$ in the
rational case.\\
{\bf Theorem 2.} {\it Let a basis $(\aaa'_1,\aaa'_2,\aaa'_3)$ of
the lattice $\Lambda$ be choosen in such a way that
$$
\eta=\frac{\BB(\aaa'_1\times\aaa'_2)}{\Phi_0}=\frac{N}{M}>0,\quad
\BB(\aaa'_2\times\aaa'_3)=\BB(\aaa'_3\times\aaa'_1)=0,
$$
where $N/M$ is the irreducible rational fraction. Then the
spectrum of the operator $H_A$ is essential and is divided into
two parts $\Sigma_1$ and $\Sigma_2$. The first one, $\Sigma_1$,
consists of infinitely degenerated Fock--Darwin levels $E_{mn_\rho
n_z}$, where $m\neq0$. The second part $\Sigma_2$ is the band
spectrum of $H_A$. This spectrum consists of bands $I_n$,
$n\geq0$, lying on the intervals $(\eps_{n},\eps_{n+1})$ (here
$\eps_{-1}=-\infty$):
$$
\Sigma_2=\bigcup\limits_{n=0}^{\infty}I_n, \quad I_n\subset(\eps_n,\eps_{n+1}).
$$
The each band $I_n$, $n\geq0$, splits into $M$ "magnetic"\ subbands $I_{n,l}$,
$l=1,2,\ldots,M$:
$$
I_n=\bigcup\limits_{l=1}^{M}I_{n,l},
$$
where $I_{n,l}=\{E_{n,l}(\pp):\pp\in\TT_3\}$, and for fixed $\pp$ the collection
$\{E_{n,1}(\pp),\ldots,E_{n,M}(\pp)\}$ is the set of all solutions of the dispersion
equation
$$
\det[q(E)+\widetilde A(\pp)]=0,
$$
lying on the interval $(\eps_n,\eps_{n+1})$.

The subbands $I_{n,l}$ of the band $I_n$ can be overlapped. Some of subbands can be
degenerated to a point} \black

Therefore there are three kinds of points in the spectrum of the
operator $H_A$. The following physical interpretation of the
described structure of the energy spectrum of our model may be
proposed. The points of the first kind (the levels of the single
quantum dot) correspond to the classical orbits lying wholly in
the single dot. The points of the second kind (the extended
states) correspond to the classical propagation trajectories in
the periodic system. Finally, the points of the third kind (the
bound states satisfying the dispersion equation) correspond to the
classical orbits lying in compact domain (cf. \cite{GPPJMP},
\cite{GP}).

As a result of numerical analysis of the dispersion equation
(\ref{DispEq}) we obtain angle--energy and flux--energy diagrams
represented on Fig.~4 and Fig.~5 respectively (we use a system of
units such that $c=\hbar=e=1$ and $m^*=1/2$). According to
Theorem~2, the bands of the spectrum look like deformed Hofstadter
butterflies (cf. \cite{GPPO}).



\section{Acknowledgements}

The work is partially supported by DFG (Grant No.~436 RUS
113/572). During the preparation of this work we have useful
discussions of mathematical aspects of the problem with
S.~Yu.~Dobrokhotov, P.~Exner, and K.~Pankrashkin, physical aspects
with M.~V.~Budantsev, V.~Ya.~Demikhovskii, D.~Grigoriev, and
M.~Semtsiv, and aspects of computer calculations with
E.~Grishanov, I.~Lobanov, and E.~Semenov. To all of these people
we express our deep gratitude.

\newpage

\centerline{\bf Figure Captions}

\vskip1cm

\noindent{\bf Fig.~1.} Three-dimensional quantum dots array and basic vectors of array lattice.\\[3mm]

\noindent{\bf Fig.~2.} The behaviour of the function $q(x)$.\\[3mm]

\noindent{\bf Fig.~3.} The behaviour of the functions $\tau_n(x)$.\\[3mm]


\noindent{\bf Fig.~4.} The angle-energy diagram for the simple-cubic array of
spherical
quantum dots.\\[3mm]

\noindent{\bf Fig.~5.} The flux-energy diagram for the simple-cubic array of
spherical
quantum dots.\\[3mm]

\newpage






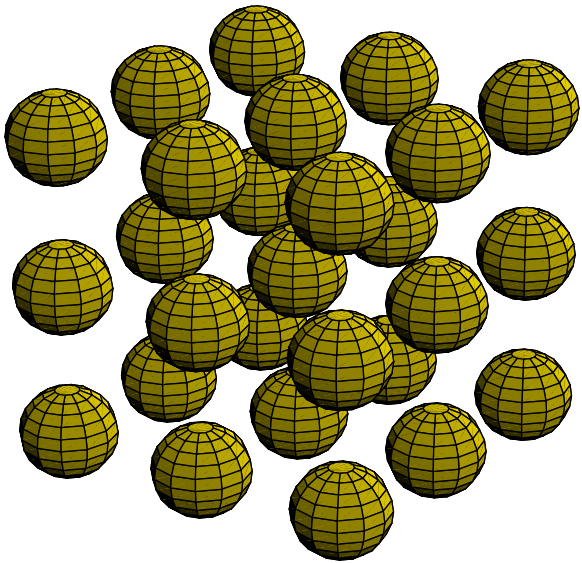
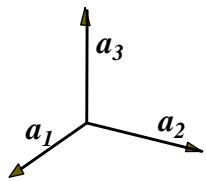
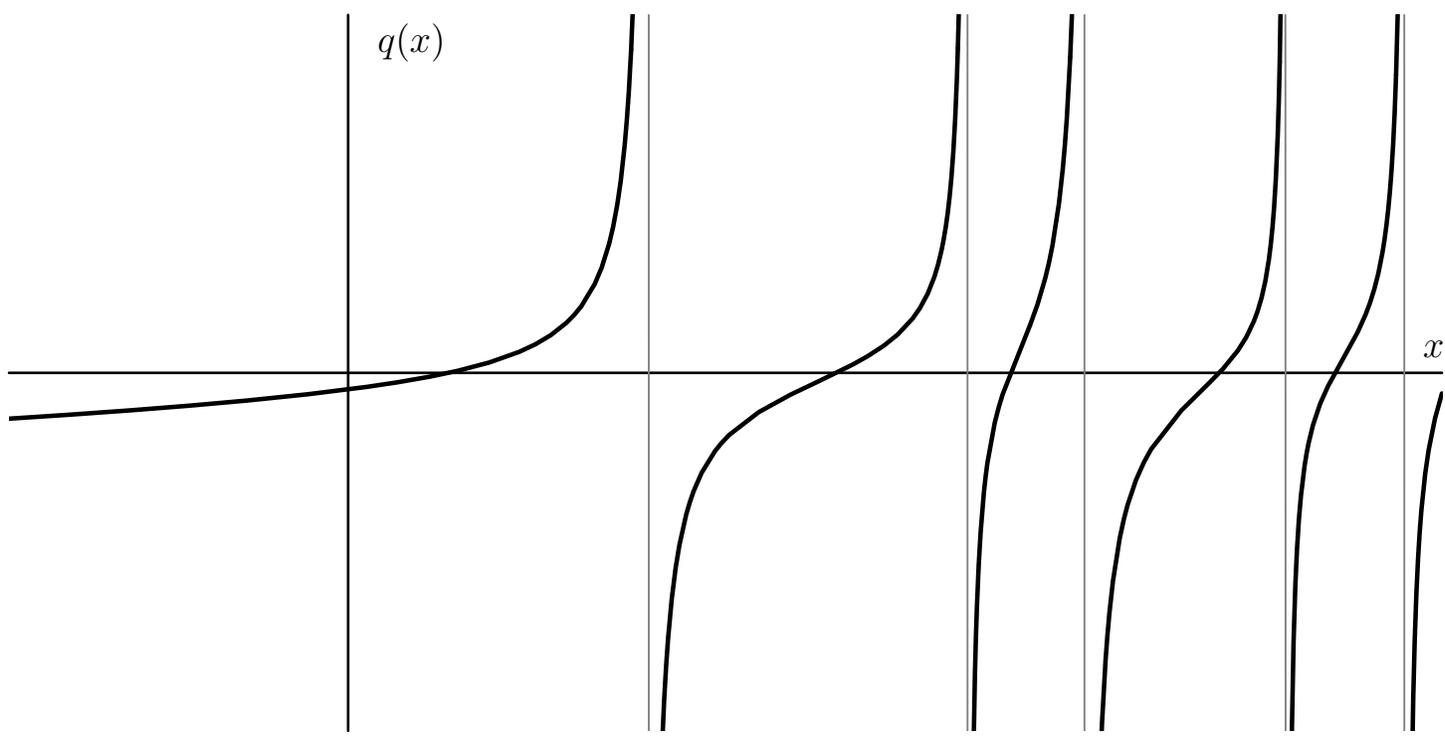
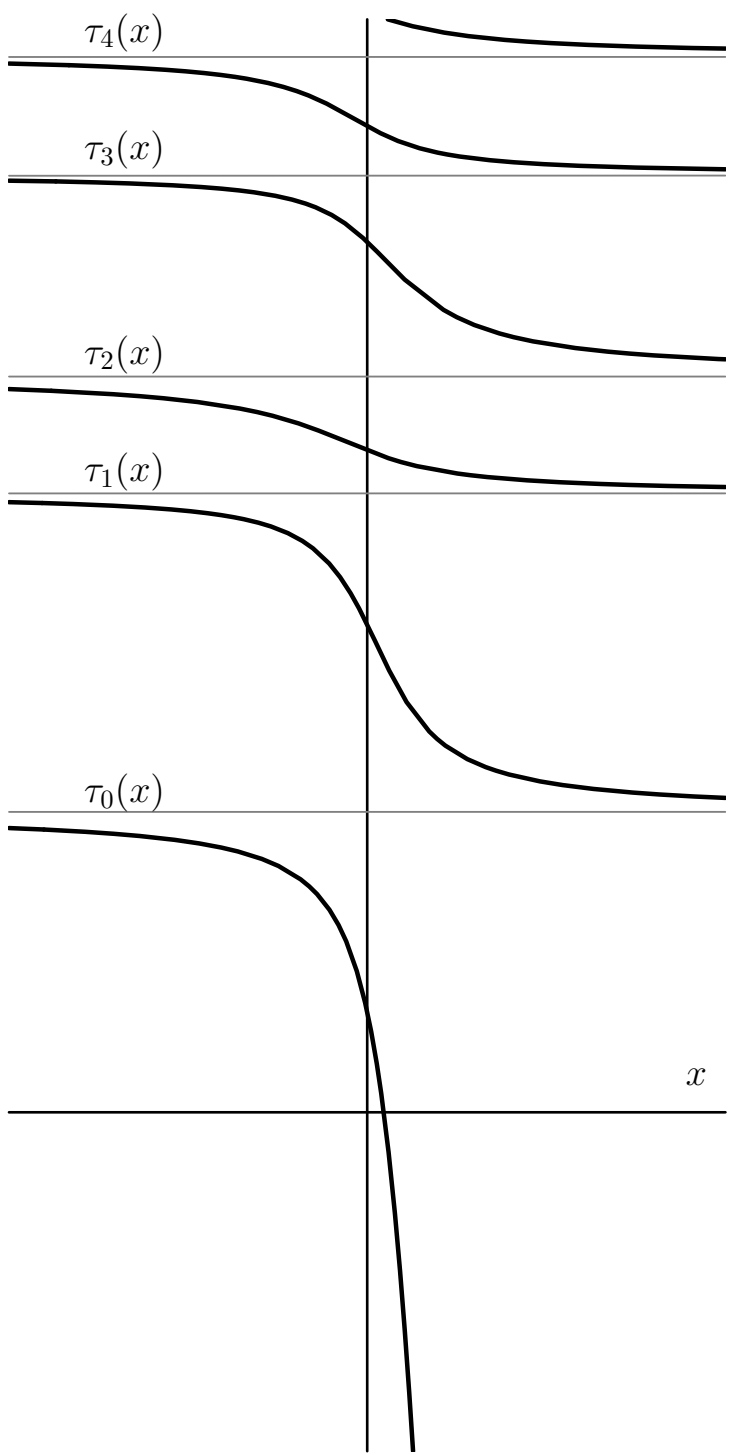
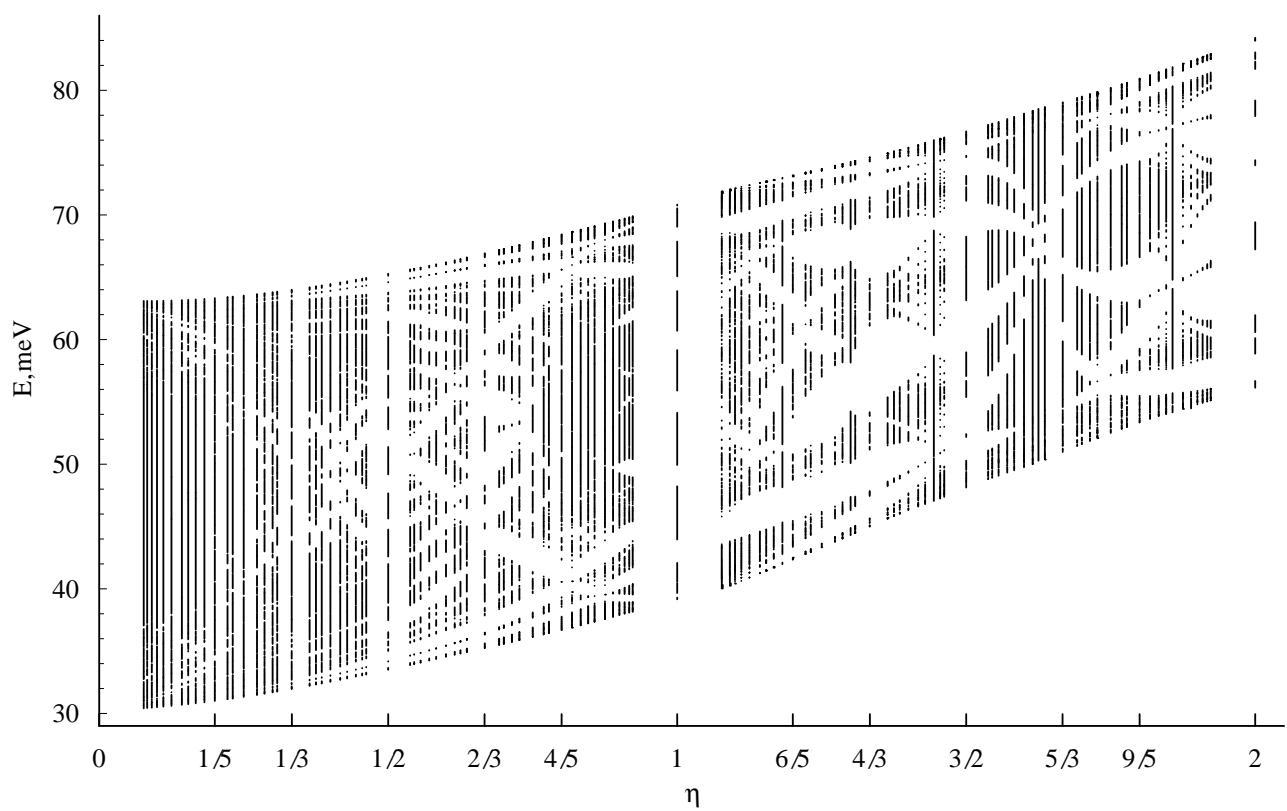
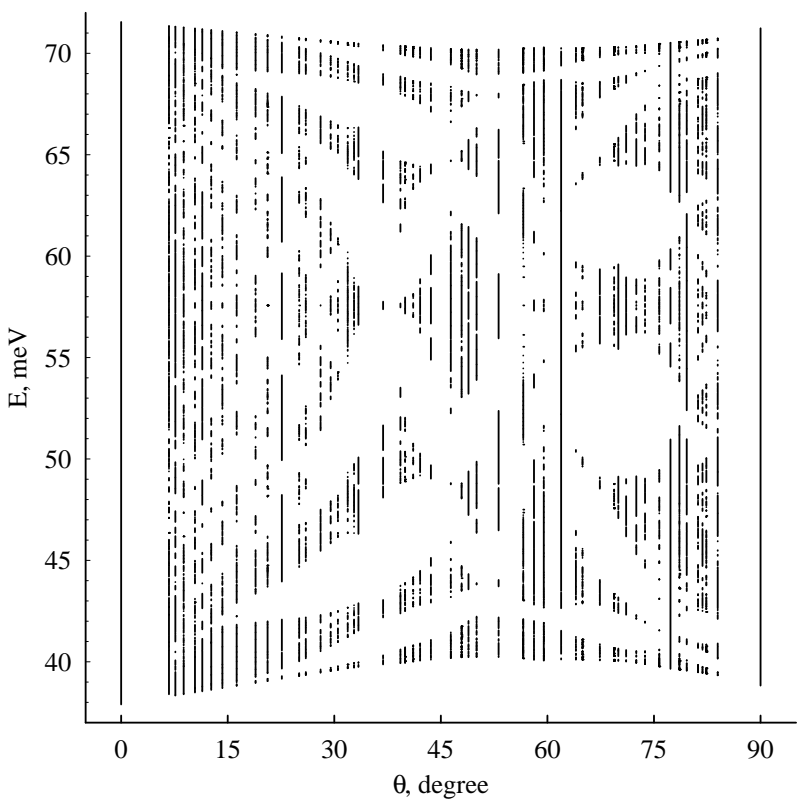
\end{document}